# Laborexperimente zur Entstehung und Wirkung kosmischer Magnetfelder

Frank Stefani

Helmholtz-Zentrum Dresden-Rossendorf
Institut für Fluiddynamik
PF 510119
01314 Dresden
E-Mail: F.Stefani@hzdr.de

## 1 Einleitung

Wer einst den Magnetismus entdeckt hat, wird sich wohl nie mehr feststellen lassen. Sicher ist, dass bereits Thales von Milet (ca. 640-546 v.u.Z.) die bemerkenswerte Eigenschaft von Magneteisenstein, Eisen anzuziehen, kannte. Auch in China wusste man um dieses Phänomen: Wahrscheinlich schon zwei Jahrhunderte v.u.Z. hatte man dort aus Magneteisenstein den ersten Kompass in Form eines Löffels hergestellt, der sich auf einer glatten Unterlage frei drehen konnte. Thales versuchte, sich dieses Verhalten mit der Beseeltheit des Magneteisensteins zu erklären, die Chinesen bezeichneten ihn als "liebenden Stein".

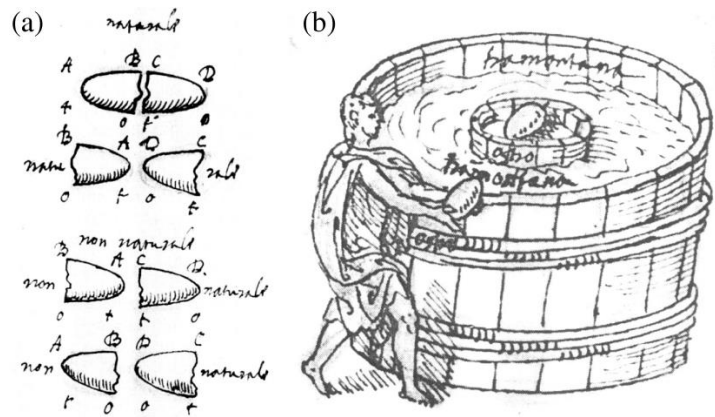

*Abb. 1: Petrus Peregrinus' Experimente mit Magneteisenstein. (a) Experimente zum Nachweis der anziehenden („naturale") und abstoßenden („non naturale") Wirkung von Magneten. (b) Ein Experiment mit einem Kompass. (Bilder aus: Petrus Peregrinus de Maricourt, Opera, Scuola normale superiore, Pisa, 1995)*

Mit Petrus Peregrinus' Abhandlung "Epistola de Magnete" aus dem Jahre 1269 begann ein neues Kapitel in der Erforschung des Magnetismus. Im Gegensatz zu manchen seiner spekulierenden Zeitgenossen ließ sich Peregrinus von den Beobachtungen leiten, die er in eigenen Experimenten mit Magneteisenstein gemacht hatte (Abb.1). Auf ihn geht der Begriff der Pole zurück; er zeigte als erster, dass sich gleichnamige Pole abstoßen und entgegengesetzte Pole anziehen (Abb. 1a). Nach seinen Experimenten zur Kompasswirkung (Abb. 1b) dauerte es aber noch mehr als dreihundert Jahre, bis William Gilbert, ebenfalls von verschiedenen

Experimenten an Kugeln aus Magneteisenstein inspiriert, die Erde selbst als großen Magneten beschreiben sollte. Die Erde - eine große Kugel aus Magneteisenstein: diese Hypothese war durch Beobachtungen gut begründet und durch Peregrinus' und Gilberts Experimente zusätzlich gestützt. Falsch war sie trotzdem...

Aus seismischen Messungen kennen wir heute die Struktur des Erdinneren recht genau. Und damit wissen wir, dass es im Kern der Erde viel zu heiß für jegliche Art von Ferromagnetismus ist. Weiterhin ist inzwischen bekannt, dass neben der Erde und anderen Planeten auch Sterne und ganze Galaxien eigene Magnetfelder haben (vgl. [1]), bei denen Ferromagnetismus als Ursache völlig ausgeschlossen werden kann.

Welche alternativen Quellen kosmischer Magnetfelder kommen aber in Betracht, wenn Ferromagnetismus ausscheidet? Bekannt ist, dass elektrische Ströme Magnetfelder erzeugen. Aber warum fließen elektrischen Ströme in Planeten, Sternen und Galaxien, und von welcher „Batterie" werden sie angetrieben? Diese Frage hatte noch Albert Einstein als eines der großen Rätsel der Physik angesehen.

Gemäß der heute akzeptierten Theorie kosmischer Magnetfelder sind es elektrisch leitfähige Fluide, z.B. flüssiges Eisen in Planetenkernen oder Plasma in Sternen, deren komplexe Strömungen Magnetfelder vermittels des sogenannten *homogenen Dynamoeffekts* hervorbringen. Da sowohl die Phänomenologie als auch die theoretischen Grundlagen kosmischer Dynamos in den voranstehenden Artikeln dieses Heftes ausführlich beschrieben worden sind, soll hier nur noch einmal das Grundprinzip veranschaulicht werden (Abb. 2a): Unter der Wirkung eines angenommenen Magnetfeldes induziert die Strömung eines leitfähigen Fluides einen elektrischen Strom, der das Magnetfeld verstärken kann. Falls der Verstärkungsfaktor unendlich wird, was eine hinreichende Stärke der Strömung und einen gewissen Grad an „Schraubenförmigkeit" voraussetzt, sprechen wir von Selbsterregung. Der Nachweis dieses Effektes in verschiedenen Flüssigmetall-Experimenten wird den ersten Schwerpunkt dieses Artikels darstellen.

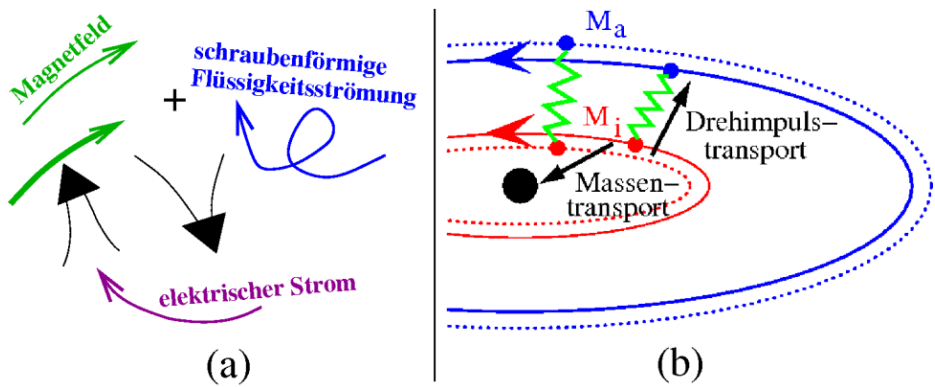

*Abb. 2: Illustration der Erzeugung und Wirkung kosmischer Magnetfelder. (a) Dynamoeffekt. (b) Magneto-Rotationsinstabilität. Magnetfelder (grün) wirken ähnlich wie elastische Federn und koppeln innere ($M_i$) und äußere ($M_a$) Massenelemente.*

Zunächst widmen wir uns aber kurz einem Problem der kosmischen Strukturbildung, für dessen Lösung magnetische Felder eine erstaunlich zentrale Rolle spielen. Planeten und Sterne drehen sich, und auch das Material, aus dem sie hervorgegangen sind, hat sich einst gedreht. Um kompakte Gebilde wie Sterne oder auch Schwarze Löcher hervorzubringen, muss der Drehimpuls zum überwiegenden Teil nach außen abtransportiert werden. Dies geschieht in sogenannten Akkretionsscheiben, die innen schneller rotieren als außen. Eine gewisse Zähigkeit vorausgesetzt, wird Drehimpuls durch Reibung nach außen transportiert. Dadurch kann drehimpulsabgereicherte Materie nach innen strömen und schließlich vom zentralen Objekt akkretiert werden. Die dabei entstehende Reibungswärme wird abgestrahlt, was z.B. die gewaltige Leuchtkraft von Quasaren erklärt. Allerdings reicht die molekulare Viskosität des Gases in Akkretionsscheiben bei Weitem nicht aus, um die beobacheteten Wachstumsraten von Sternen und Schwarzen Löchern zu erklären. Dieses Problem kann nur durch ein ausreichendes Maß an Turbulenz gelöst werden. Die Enstehungsursache dieser Turbulenz ist nun aber eines der großen Rätsel der astrophysikalischen Hydrodynamik. Nach einem Kriterium, das vor über 120 Jahren Lord Rayleigh aufgestellt hatte, sollten Strömungen in Akkretionsscheiben laminar sein und bei kleinen Störungen keinen Übergang zur Turbulenz zeigen.

Verglichen mit der Schwerkraft spielen magnetische Kräfte im Allgemeinen eine untergeordnete Rolle. Für das Funktionieren von Akkretionsscheiben sind aber gerade sie ausschlaggebend. Seit einem halben Jahrhundert ist bekannt, dass magnetische Kräfte in der Lage sind, in einer stabilen Rotationsströmung Turbulenz zu entfachen. Diese sogenannte Magneto-Rotationsinstabilität (MRI) war bereits 1959 von Jewgeni Welichow entdeckt, in ihrer Bedeutung für die Astrophysik aber erst 1991 von Steven Balbus und John Hawley erkannt worden. Die MRI bewirkt, dass hydrodynamisch stabile Rotationsströmungen, inklusive der Keplerströmungen um kosmische Zentralobjekte, durch externe Magnetfelder destabilisiert und dadurch turbulent werden können (Abb. 2b). Erst durch die MRI wird der notwendige Drehimpulstransport innerhalb von Akkretionsscheiben ermöglicht, ohne den die beobachtete Massenkonzentration in Sternen und Schwarzen Löchern nicht zu verstehen wäre.

Erklärt der homogene Dynamoeffekt, wie kosmische Magnetfelder entstehen, so beschreibt die MRI, wie Magnetfelder auf die Strukturbildung im Kosmos zurückwirken. Beide Effekte werden seit Jahrzehnten mit theoretisch-numerischen Methoden intensiv erforscht. Aber auch auf diesem Gebiet der Geo- und Astrophysik sollte man die Warnung beherzigen, die Richard Feynman auf seiner Wandtafel hinterlassen hatte: „What I cannot create I do not understand." Wie homogene Dynamos und magnetisch getriggerte Strömungsinstabilitäten funktionieren, und zwar nicht nur auf dem Papier oder im Computer, sondern in konkreten Experimenten, das ist das zentrale Thema dieses Artikels (vgl. [2]).

**2 Das Prinzip der Selbsterregung**

Die Entdeckung des Dynamoprinzips wird gewöhnlich Werner von Siemens zugeschrieben. Offenbar aber lag diese Erfindung in der Mitte des 19. Jahrhunderts in der Luft, da auch Zeitgenossen wie Hjort, Jedlik und Wheatstone sich intensiv mit der Magnetfeld-Selbsterregung beschäftigt hatten.

An Hand eines Gedankenexperimentes lässt sich das Prinzip der Selbsterregung einfach veranschaulichen. Im sogenannten Scheibendynamo (Abb. 3a) ist eine metallische Scheibe auf einer Achse drehbar gelagert. Scheibe und Achse sind über zwei Schleifkontakte und einen

Draht miteinander verbunden, wobei der Draht in einer Windung um die Achse herumgeführt wird. Dreht man die Scheibe in einem angenommenen äußeren Magnetfeld $B_0$, so wird eine elektromotorische Kraft induziert, die proportional zum Produkt aus Drehzahl und Magnetfeld ist. Diese treibt einen elektrischen Strom durch den Draht, der seinerseits ein zusätzliches Magnetfeld $B_1$ erzeugt, welches parallel zum ursprünglich angelegten Feld $B_0$ gerichtet ist. Mit anderen Worten: das ursprüngliche Feld wird verstärkt, und zwar umso mehr, je höher die Drehzahl und die gegenseitige Induktivität L zwischen Draht und Scheibe, und je niedriger der elektrische Widerstand R des Stromkreises ist. Bei einer bestimmten kritischen Drehzahl wird die Verstärkung, also das Verhältnis aus Gesamtmagnetfeld $B=B_0+B_1$ zu angelegtem Magnetfeld $B_0$, unendlich (Abb. 3b); dies ist der Punkt, an dem die Selbsterregung eines Magnetfeldes einsetzt.

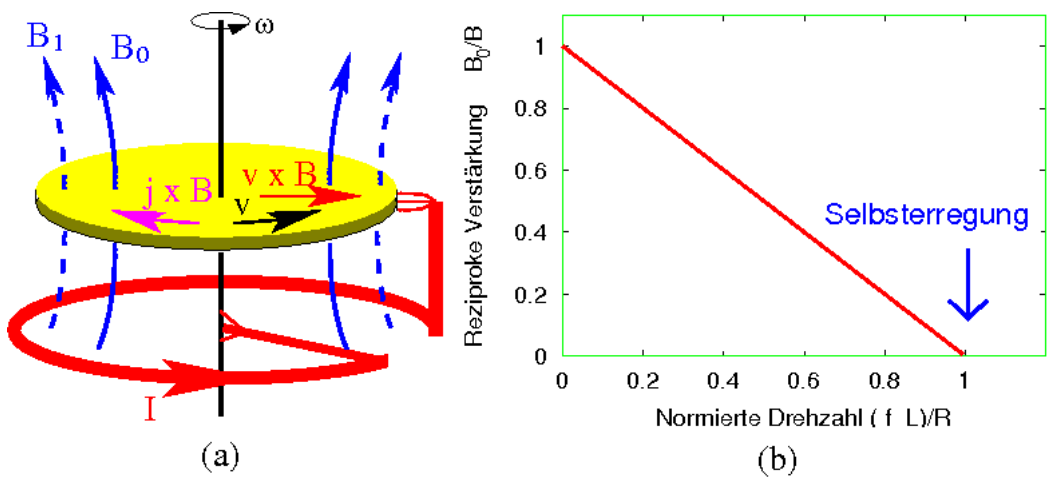

*Abb. 3: Prinzip des Scheibendynamos: (a) Eine metallische Scheibe dreht sich mit Drehzahl f in einem äußeren Magnetfeld $B_0$. Über Schleifkontakte am Scheibenrand und an der Achse wird ein Strom abgegriffen und so um die Achse geführt, dass das induzierte Magnetfeld $B_1$ das angelegte Magnetfeld $B_0$ verstärkt. (b) Reziproker Wert $B_0/B$ der Verstärkung des Scheibendynamos in Abhängigkeit von der Drehzahl f. Im stationären Fall stellt sich ein Gesamtfeld $B=B_0+B_1=B_0/(1-f L/R)$ ein, wobei L die gegenseitige Induktivität zwischen Windung und Scheibe und R den Gesamtwiderstand des Stromkreises bezeichnet. Bei der kritischen Drehzahl f=R/L wird die Verstärkung formal unendlich; an diesem Punkt tritt Selbsterregung ein.*

Wird die Drehzahl über diesen kritischen Punkt hinaus erhöht, beginnt das Magnetfeld exponentiell zu wachsen. Bei festgehaltener Drehzahl würde dieses exponentielle Wachstum unbegrenzt fortschreiten. Tatsächlich aber gibt es eine Grenze des Wachstums, die durch die Rückwirkung der erzeugten Magnetfelder und Ströme auf die Rotation der Scheibe gesetzt wird. Diese seiner Ursache entgegen gerichtete Wirkung induzierter Ströme und Felder ist als Lenz'sche Regel bekannt. Schließlich wird sich ein Gleichgewichtszustand einstellen, in dem das die Scheibe antreibende Drehmoment gerade durch das rückwirkende Drehmoment kompensiert wird. Diese Phase wird als Sättigungsphase bezeichnet, während die vorangehende Phase exponentiellen Wachstums, in der das Magnetfeld zum Bremsen der Scheibe noch zu klein ist, als kinematische Phase bezeichnet werden kann.

Verstärkung eines angelegten Feldes, Selbsterregung, Sättigung: all dies sind Erscheinungen, die man auch bei einem homogenen Dynamo wiederfindet. Allerdings ist es keineswegs selbstverständlich, wie sich der Rückkopplungsmechanismus des Magnetfeldes von einer technischen Einrichtung wie dem Scheibendynamo auf eine homogene Flüssigkeit übertragen läßt. Schließlich sind die Zwänge, die dem Stromfluß durch die geometrische Anordnung des Scheibendynamos (Scheibe, Draht, Achse) auferlegt werden, in einer Flüssigkeit nicht vorhanden.

Ganz analog zum Scheibendynamo gibt es aber auch in homogenen Dynamos einen Wettbewerb zwischen Magnetfeldverstärkung und Ohmschen Verlusten. Durch den Vergleich beider Terme kann man zeigen, dass es erst dann zur Selbsterregung kommt, wenn die magnetische Reynoldszahl Rm:=$\mu_0$ $\sigma$ L V einen kritischen Wert übersteigt. Dabei ist $\mu_0$ die magnetische Permeabilitätskonstante, $\sigma$ die elektrische Leitfähigkeit, L eine typische Längenausdehnung des Fluides und V eine typische Geschwindigkeit der Strömung. Es gehört zum Wesen der Selbsterregung, dass dem Dynamo ein beliebig kleines und beliebig geformtes Anfangsfeld genügt, um daraus das für ihn charakteristische Magnetfeld zu generieren.

Will man konkrete kosmische oder experimentelle Dynamos untersuchen, so kommt man um eine genauere mathematische Formulierung des eben skizzierten Prozesses nicht herum. Die zeitliche Entwicklung des Magnetfeldes wird durch die sogenannte Induktionsgleichung beschrieben, deren Lösung – abgesehen von ganz wenigen einfachen Modellen – nur numerisch möglich ist. Diese Computersimulationen sind in den letzten 50 Jahren enorm verfeinert worden. Inzwischen umfassen sie auch die simultane Lösung der Navier-Stokes-Gleichungen, mit denen die Flüssigkeitsbewegung – auch unter Einbeziehung der Rückwirkung des Magnetfeldes auf die Strömung - beschrieben wird. Auf diese Weise wurden z.B. sehr beeindruckende Berechnungen für den Geodynamo durchgeführt, in denen sogar die sporadischen Magnetfeldumpolungen simuliert werden konnten. Allerdings reichen die derzeitigen Computerleistungen bei weitem nicht aus, um die Prozesse im Erdkern realistisch zu berechnen.

**3 Dynamoexperimente**

Wenn es kosmische Magnetfelder in solchem Überfluss gibt, worin besteht dann das Problem, den homogenen Dynamoeffekt im Labor nachzubilden? Wir hatten bereits die magnetische Reynoldszahl Rm als diejenige dimensionslose Kennzahl identifiziert, welche über die Möglichkeit der Erzeugung von Magnetfeldern in leitfähigen Flüssigkeiten entscheidet. Was aber bedeutet es für eine Strömung konkret, wenn man einen typischen Wert des kritischen Rm von etwa 100 erreichen muss? Die elektrische Leitfähigkeit des besten flüssigen Leiters, Natrium, beträgt etwa $10^7$ $(\Omega$ m$)^{-1}$. Für ein Laborexperiment in einer Kugel von 1 m Radius wäre dann eine Geschwindigkeit von fast 10 m/s erforderlich. Der Aufbau eines Dynamoexperimentes mit diesen Größenordnungen von Länge und Geschwindigkeit ist ein aufwendiges Unterfangen, vor allem in Anbetracht der notwendigen Sicherheitsvorkehrungen beim Umgang mit großen Mengen flüssigen Natriums. Kosmische Körper hingegen erreichen infolge ihrer großen Längenskala problemlos ausreichende Werte von Rm, selbst bei geringeren Leitfähigkeiten und kleineren Strömungsgeschwindigkeiten der Fluide.

**3.1 Das Rigaer Dynamoexperiment**

Im Rigaer Dynamoexperiment (siehe Abb. 4), welches unter Leitung von Agris Gailitis seit Mitte der 70er Jahre konzipiert und seit Anfang der 90er des vorigen Jahrhunderts Jahre aufgebaut worden war, soll weder der flüssige Erdkern, noch unsere Sonne, noch eine Galaxie nachgebildet werden. Vielmehr wird die Selbsterregung eines Magnetfeldes in einer „Elementarzelle" homogener Dynamos, einer schraubenförmigen Flüssigkeitsströmung, untersucht. Abbildung 4 zeigt das Herzstück des Dynamoversuchsstands: eine koaxiale Anordnung von drei mit flüssigem Natrium gefüllten Edelstahlröhren von etwa 3 m Länge und einem Außendurchmesser von 0.8 m. Im inneren Rohr wird das Natrium mittels eines Propellers schraubenförmig nach unten gepumpt und dort umgelenkt. Danach strömt es im zweiten Rohr geradlinig nach oben, wo es wiederum in die Propellerregion zurückgelangt. Der äußere Zylinder enthält stehendes Natrium. Die Geschwindigkeiten des Natriums erreichen Werte von bis zu 20 m/s, wofür ein Propellerantrieb mit einer Gesamtleistung von etwa 200 kW notwendig ist. Der damit erzielbare Wert von Rm liegt etwa bei 25.

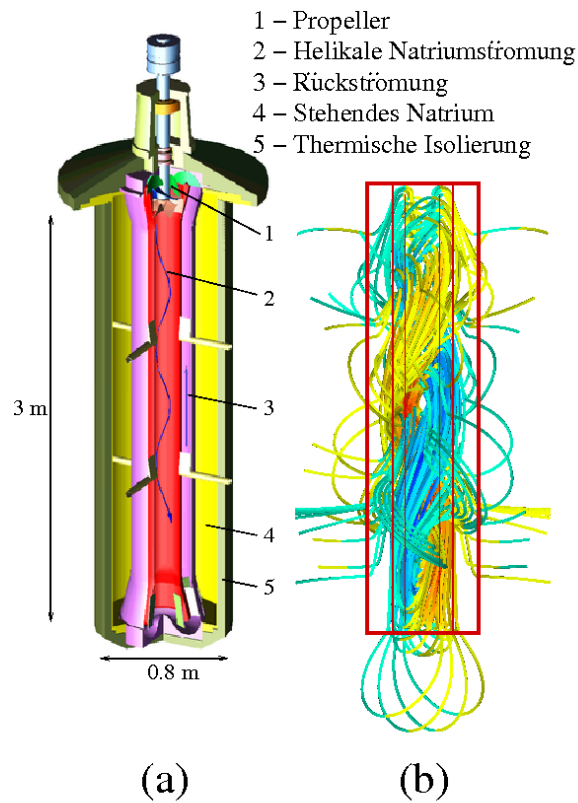

Abb. 4.: Das Rigaer Dynamoexperiment. (a) Zentraler Modul. (b) Berechnete Struktur des selbsterregten Magnetfeldes.

Was hier recht einfach klingt, basiert auf jahrzehntelangen theoretischen Vorarbeiten. Zunächst hatte Ponomarenko 1973 theoretisch nachgewiesen, dass die schraubenförmige Bewegung eines elektrisch leitfähigen, unendlich langen Zylinders in einem gleichfalls leitfähigen, aber feststehendem Medium eine konvektive magnetische Instabilität erzeugen kann. Dies bedeutet, daß ein exponentielles Wachstum eines Magnetfeldes zwar auftritt, dieses sich aber mit dem Zylinder mitbewegt, so daß an einem festgehaltenen Punkt kein exponentielles Wachstum messbar wäre. Agris Gailitis' Idee war es, durch die Rückströmung in einem zweiten koaxialen Zylinder eine Rückkopplung des im inneren Zylinder angefachten Magnetfeldes auszulösen. Effektiv bildet sich dabei eine stehende Magnetfeldwelle aus, so dass exponentielles Wachstum tatsächlich im Laborsystem beobachtbar wird. In Kooperation mit dem Helmholtz-Zentrum Dresden-Rossendorf (HZDR) und der TU Dresden wurde dann die Strömungsstruktur im Innenzylinder so lange optimiert, bis die Selbsterregung eines Magnetfeldes in Form einer Doppelhelix (Abb. 4b) als hinreichend wahrscheinlich angesehen werden konnte.

Im November 1999 war es schließlich soweit: das erste Experiment mit flüssigem Natrium konnte starten. Für den Scheibendynamo hatten wir gezeigt, dass vor der eigentlichen Selbsterregung zunächst die Verstärkung eines angelegten Feldes auftritt. Diese Verstärkung zu untersuchen, war das erste Ziel des Experiments. Zu diesem Zweck wurde mittels einer um die Anlage herum gewickelten Erregerspule ein 1-Hz Wechselfeld angelegt und dessen Verstärkung durch den Dynamo gemessen. Bei der höchsten Propellerdrehzahl von 2150/min tauchte zusätzlich zum verstärkten 1-Hz Signal plötzlich ein weiteres Signal mit einer Frequenz von etwa 1.3 Hz auf, welches exponentiell anwuchs. Damit war die Selbsterregung eines Magnetfeldes in einer Flüssigmetallströmung erstmalig nachgewiesen worden, bevor das Experiment wegen eines kleinen Natrium-Lecks abgebrochen werden musste.

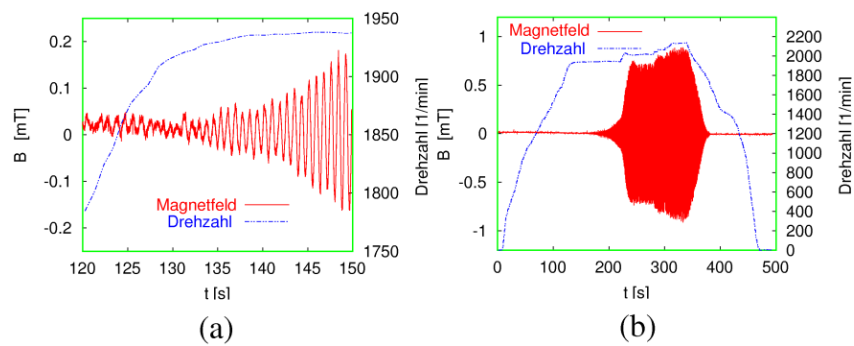

*Abb. 5: Ein typischer Lauf am Rigaer Dynamoexperiment: (a) Propeller-Rotationsrate und Magnetfeldmessung im Inneren des Dynamos. Bei einer Rotationsrate von etwa 1900/min setzt Selbsterregung ein. Ein Magnetfeld in der Form wie in Abb. 4b beginnt exponentiell zu wachsen und um die vertikale Achse zu rotieren, wodurch der stehende Magnetfeldsensor ein oszillierendes Magnetfeld sieht. (b) Rotationsrate und Magnetfeldmessung außerhalb des Dynamos für die gesamte Laufzeit. Nach einer Phase exponentiellen Wachstums stellen sich – in Abhängigkeit von der Rotationsrate – verschiedene Sättigungszustände ein.*

Nach der notwendigen Reparatur wurde im Juli 2000 eine zweite Serie von Experimenten durchgeführt. Diesmal war es möglich, bei tieferen Temperaturen und somit höherer Leitfähigkeit des Natriums zu arbeiten, was das Einsetzen der Selbsterregung schon bei niedrigeren

Drehzahlen gestattete. Das Ergebnis eines dieser experimentellen Läufe ist in Abb. 5 dokumentiert. Die blaue Kurve zeigt die Drehzahl des Propellers, die rote Kurve das gemessene Magnetfeld. In Abb. 5a sieht man deutlich, wie sich bei einer Drehzahl von etwa 1900/min ein oszillierendes Signal aus dem Rauschen heraus entwickelt und exponentiell zu wachsen beginnt. Das Signal eines anderen, außen angebrachten Magnetfeldsensors illustriert den Gesamtverlauf dieses Experiments, bestehend aus einer Anfangsphase mit vernachlässigbarem Magnetfeld, der kinematischen Phase mit exponentiellem Wachstums, einer Sättigungsphase sowie einer Abklingphase nach dem Herunterfahren der Drehzahl. Aus einer Vielzahl von Experimenten dieser Art konnten detaillierte Abhängigkeiten der Wachstumsrate, der Frequenz sowie der räumlichen Struktur des selbsterregten Magnetfeldes gewonnen werden. Dabei wurde jeweils eine gute Übereinstimmung mit den numerischen Prognosen festgestellt.

### 3.2 Das Karlsruher Dynamoexperiement

Fast gleichzeitig mit dem Rigaer Experiment ging Ende 1999 ein zweites großes Dynamo-Experiment in Betrieb. Diese Koinzidenz ist umso bemerkenswerter, als die Konzeption beider Experimente Jahrzehnte und ihr eigentlicher Aufbau immerhin noch mehrere Jahre in Anspruch genommen hatte. Interessant ist auch die wissenschaftliche Komplementarität beider Experimente: Während im Rigaer Experiment eine einzelne großskalige schraubenförmige Strömung ein Magnetfeld von vergleichbarer räumlicher Ausdehnung erzeugt, generieren die 52 kleinskaligen Spin-Erzeuger im Karlsruher Experiment (Abb. 6a) gemeinsam ein großskaliges Magnetfeld von der Ausdehnung des Gesamtsystems. Diese Trennung zwischen kleinskaliger Strömungsstruktur und resultierender Dynamowirkung auf einer größeren Skala ist eine nahezu perfekte Illustration des mean-field-Dynamokonzepts, das in den 60er Jahren von Max Steenbeck, Fritz Krause und Karl-Heinz Rädler entwickelt worden war. Konkret lässt sich die kollektive Induktionswirkung der vielen Spin-Erzeuger als sogenannter $\alpha$-Effekt beschreiben, der bewirkt, dass eine elektromotorische Kraft parallel zu einem vorhandenen Magnetfeld erzeugt wird. Auf Grundlage dieses Konzeptes konnte das Verhalten des Karlsruher Dynamos durch die Gruppe um Karl-Heinz Rädler numerisch sehr detailliert vorhergesagt worden.

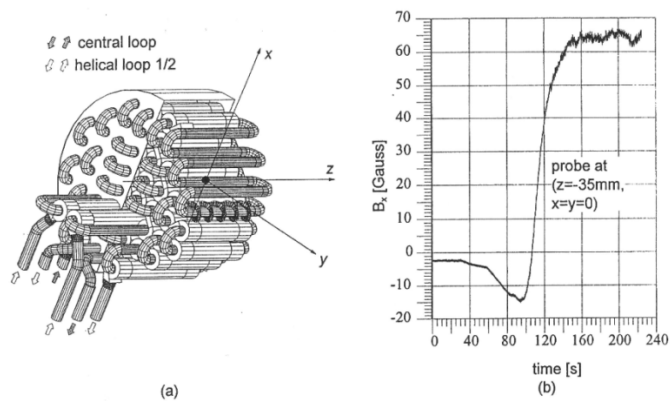

Abb. 6: Das Karlsruher Dynamoexperiment: (a) Aufbau des zentralen Moduls: Mittels externer elektromagnetischer Pumpen wird Natrium durch die zentralen und die schraubenförmigen Kanäle von 52 Spin-Erzeugern gepumpt. (b) Ab einer kritischen Pumpleistung kommt es zur Selbsterregung eines Magnetfeldes, welches an einem Sensor im Dynamomodul gemessen wird (Quelle: Robert Stieglitz, KIT).

Abbildung 6b demonstriert beispielhaft, wie ein Magnetfeld im Karlsruher Experiment erzeugt wird. Das gezeigte Signal wurde bei fest eingestellter Durchflussrate von 115 m$^3$/h in den zentralen Kanälen der Spin-Erzeuger gemessen, nachdem die Durchflussrate in den helikalen Kanälen bei t=30 s von 95 m$^3$/h auf 107 m$^3$/h erhöht worden war. Der sehr steile Feldanstieg, der bei etwa t=100 s beginnt, ist ein klares Indiz für den Dynamoeffekt.

**3.3 Das VKS-Experiment**

Im französischen Cadarache wird seit Ende der 90er Jahre ebenfalls an einem Dynamoexperiment mit flüssigem Natrium gearbeitet. Dieses wird inzwischen als von-Kármán-Sodium-Experiment, abgekürzt VKS-Experiment, manchmal auch als „Französische Waschmaschine" bezeichnet. Letztere Bezeichnung lässt sich anhand der Abbildung 7a leicht nachvollziehen: zwei Propeller mit gekrümmten Flügeln erzeugen eine Natriumströmung, die im Wesentlichen aus zwei gegenläufig rotierenden Wirbeln sowie zwei zugehörigen poloidalen Strömungen besteht. Nach vielen vergeblichen Versuche führte die Ersetzung des üblichen nichtmagnetischen Stahls durch magnetischen Stahl schließlich zum Erfolg: 2006 konnte im VKS-Dynamo ebenfalls Selbsterregung nachgewiesen werden. Dabei gab es allerdings zwei Überraschungen: zum ersten „zündete" der Dynamo schon bei etwa zwei Drittel der numerisch prognostizierten Propellerdrehzahl. Zum zweiten war das entstehende Magnetfeld weitgehend axial-symmetrisch, im Gegensatz zu allen numerischen Rechnungen, die ein nicht-axialsymmetrisches Magnetfeld vorausgesagt hatten. Inzwischen konnten beide Effekte durch numerische Simulationen am HZDR zumindestens qualitativ erklärt werden: die magnetischen Eigenschaften des Propellermaterials erwiesen sich dabei als entscheidend für den Erfolg des VKS-Experiments.

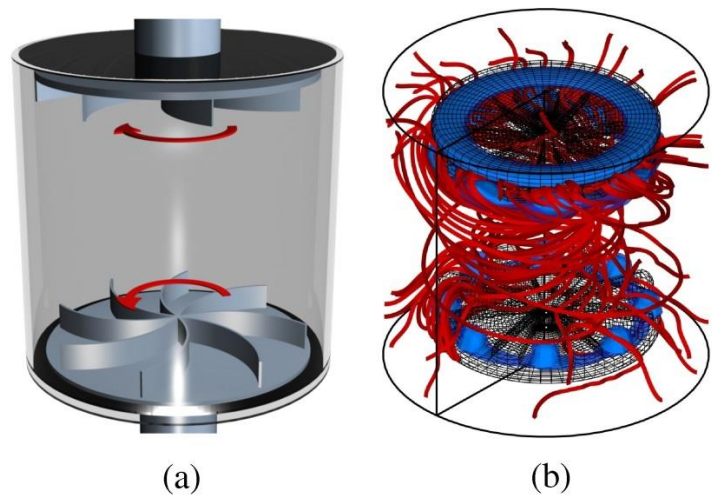

*Abb. 7: Das VKS-Experiment im französischen Cadarache. (a) Prinzipskizze des Strömungsantriebs. (b) Simulation des entstehenden Magnetfeldes unter Berücksichtigung des magnetischen Propellermaterials. (Quelle: André Giesecke, HZDR)*

Abgesehen von diesem „Schönheitsfehler" hat das VKS-Experiemnt hochinteresannte Ergebnisse erbracht. Insbesondere wurden für den Fall unterschiedlicher Rotationsraten der beiden Propeller spontane Umpolungen gefunden, wie sie auch für das Erdmagnetfeld bekannt sind.

### 3.4 Weitere Experimente

Inzwischen gibt es ein weltweites Netzwerk von sehr kollegial kooperierenden Forschungsgruppen, die an verschiedenen Flüssigmetall-Experimenten arbeiten. Auch wenn die eigentliche Dynamowirkung bisher nur in Riga, Karlsruhe und Cadarache nachgewiesen worden ist, sind andernorts ebenfalls bemerkenswerte Resultate erzielt worden. In zwei torusförmigen Strömungsexperimenten in Perm konnte z.B. der sehr wichtige $\beta$-Effekt, welcher den turbulenzbedingten Abfall der effektiven elektrischen Leitfähigkeit eines Fluids beschreibt, nachgewiesen und quantifiziert werden. Vergleichbare Resultate wurden auch in einem kugelförmigen Natrium-Experiment in Madison gewonnen. Diese Ergebnisse spielen für das bessere Verständnis planetarer Dynamos eine wichtige Rolle. In sphärischen Couette-Experimenten (d.h. in Strömungen zwischen zwei konzentrischen, differenziell rotierenden Kugeln) in Maryland und Grenoble konnten verschiedene Arten von Wellen, die unter dem gemeinsamen Einfluss von Magnetfeldern und Rotation entstehen, detailliert untersucht werden. In einem Tayler-Couette-Experiment (d.h. in einer Strömung zwischen zwei konzentrischen, differenziell rotierenden Zylindern) in Socorro konnte die etwa 8–fache Verstärkung eines angelegten Magnetfeldes gemessen werden.

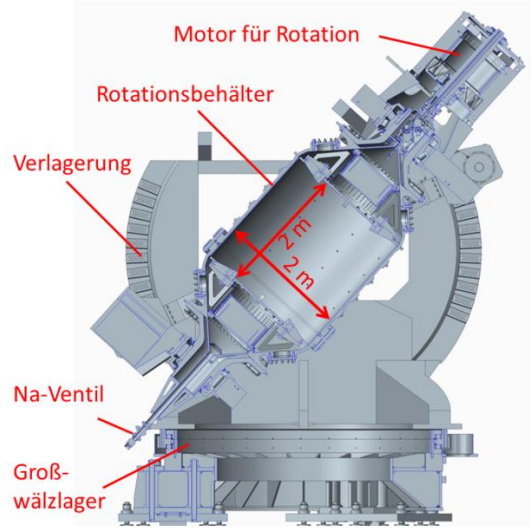

Abb. 8: Das geplante Präzessionsexperiment im Rahmen des DRESDYN-Projektes am Helmholtz-Zentrum Dresden-Rossendorf. Ein Rotationsbehälter mit etwa 8 t flüssigem Natrium rotiert um seine Längsachse mit bis zu 600/min und um eine dazu geneigte Achse mit bis zu 60/min. Die entstehenden Kreiselmomente erfordern ein extrem stabiles Fundament. (Quelle: SBS Bühnentechnik GmbH).

Ein weiteres großes Dynamoexperiment wird derzeit im Rahmen des DRESDYN-Projektes am HZDR vorbereitet. Ziel dieses Experimentes ist es zu klären, ob und unter welchen Bedingungen Präzession, d.h. die Rotation eines Körpers um zwei verschiedene Achsen, als Energiequelle für Selbsterregung in Frage kommt. Diese Problematik ist sowohl für den Geodynamo als auch z.B. für Dynamos in Asteroiden relevant. Abbildung 9 zeigt das Design dieses geplanten Experimentes.

Ein ganz neuer Typ von Dynamoexperimenten, in denen das flüssige Metall durch ein Plasma ersetzt wird, wird derzeit in Madison getestet. Der große Vorteil dieses Konzepts besteht darin, dass in Abhängigkeit von der Temperatur und der Dichte verschiedene Leitfähigkeiten und Viskositäten des Plasmas eingestellt werden können, wodurch eine große Variabilität der Parameterbereiche erreicht wird.

## 4. Experimente zur Magneto-Rotationsinstabilität

Aufgrund ihrer enormen astrophysikalischen Bedeutung gibt es seit etwa 10 Jahren weltweit intensive Bemühungen, auch die MRI im Labor zu untersuchen. Allerdings ist dies nur mit einem hohen experimentellen Aufwand möglich, sofern man – in der Tradition von WelikhoW, Balbus und Hawley – von einem rein vertikal angelegten Magnetfeld ausgeht. Ein erster Versuch in diese Richtung war 2004 mit einem sphärischen Couette-Experiment an der University of Maryland unternommen worden. In diesem konnte bei hohen Rotationsraten und Magnetfeldstärken eine interessante neue Strömungsstruktur nachgewiesen werden, deren konkrete Parameterabhängigkeit durchaus auf einen Zusammenhang mit der MRI hindeutete. Da diese neue Struktur aber auf dem Hintergrund einer bereits hoch-turbulenten Strömung auftrat, konnte der ursprüngliche Anspruch, MRI als erste Instabilität einer hydrodynamisch stabilen Strömung zu identifizieren, nicht eingelöst werden.

Ein Taylor-Couette-Experiment mit hohen Rotationsraten und starkem vertikalem Magnetfeld wird derzeit in Princeton betrieben. Hier wurde auch bereits eine Vorstufe der MRI in Form sogenannter Magneto-Corioliswellen gemessen, während der eigentliche Nachweis der Standardform der MRI noch aussteht.

Das PROMISE-Experiment am HZDR basiert auf der Entdeckung von Günther Rüdiger (AIP Potsdam) und Rainer Hollerbach (University of Leeds), dass der experimentelle Aufwand zum Nachweis der MRI drastisch reduziert werden kann, wenn das rein vertikale Magnetfeld durch ein azimutales Magnetfeld ergänzt wird. Für diese inzwischen als *helikale MRI* (HMRI) bezeichneten Variante liegt der Wert der kritischen Reynoldszahl bei etwa $10^3$ (verglichen mit etwa $10^6$ für die Standard-MRI), wodurch sich die Möglichkeit ergab, MRI tatsächlich als erste Instabilität einer hydrodynamisch stabilen Strömung nachzuweisen. Ein Schönheitsfehler der HMRI besteht allerdings darin, dass sie nur in der Lage ist, verhältnismäßig steil nach außen abfallende Rotationsprofile zu destabilisieren. Ob sie auch für die astrophysikalisch bedeutsamen Keplerströmungen funktioniert, ist Gegenstand der aktuellen Forschung.

Das PROMISE-Experiment besteht im Kern aus zwei gleichsinnig rotierenden Kupferzylindern, zwischen denen sich die Flüssigmetall-Legierung GaInSn befindet (Abb. 9a). Deren Rotationsströmung ist hydrodynamisch stabil, solange das Verhältnis der

Winkelgeschwindigkeiten von Außen- und Innenzylinder größer ist als das quadrierte Verhältnis von Innen- zu Außenradius. Erst mit angelegten vertikalen und azimutalen Magnetfeldern setzt die HMRI in Form einer nach unten oder oben wandernden Welle ein, die mit Hilfe von zwei Ultraschall-Geschwindigkeitssensoren ausgemessen werden kann.

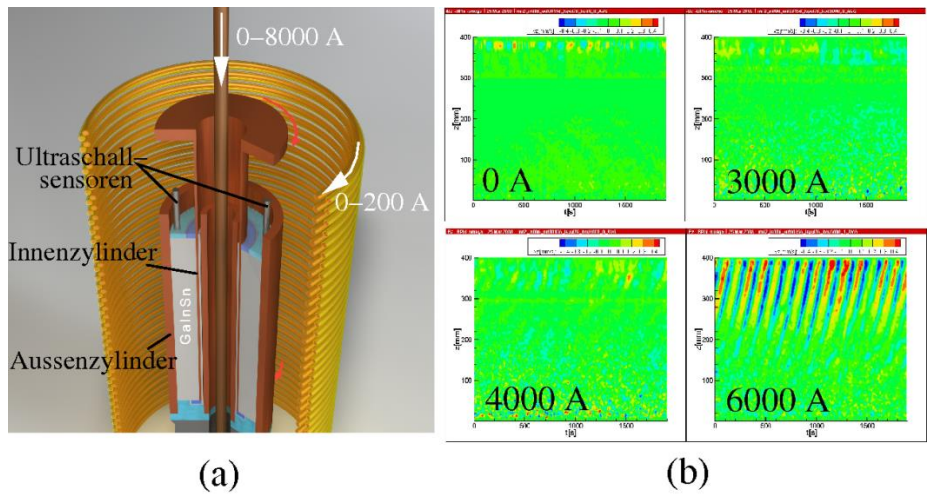

*Abb. 9: Das PROMISE-Experiment am Helmholtz-Zentrum Dresden-Rossendorf zur Untersuchung der Magneto-Rotationsinstabilität (MRI). (a) Aufbau des Experiments. Der Zylinderspalt hat eine Breite von 4 cm und eine Höhe von 40 cm. (b) Mittels Ultraschall-Sensoren gemessene Vertikalgeschwindigkeit im Flüssigmetall GaInSn in Abhängigkeit von Zeit und Höhe. Bei angelegtem Spulenstrom von 76 A tritt die MRI in Form einer wandernden Welle erst ab einem zentralen Strom von etwa 5 kA auf.*

In einer Vielzahl von Experimenten wurden verschiedene Parameter variiert und die resultierenden wandernden Wellen hinsichtlich Frequenz und Wellenzahl ausgewertet. Abbildung 9b illustriert einen typischen Experimentverlauf: Bei festgehaltenem Vertikalfeld, erzeugt durch einen Spulenstrom von 76 A, wird die Stärke des Azimutalfeldes durch Anlegen unterschiedlicher axialer Ströme im zentralen Kupferstab variiert. In guter Übereinstimmung mit numerischen Vorhersagen tritt die HMRI erst für hohe axiale Ströme ab etwa 5 kA auf.

Eine weitere Variante der MRI ergibt sich, wenn man auf das vertikale Magnetfeld komplett verzichtet. Unter dem Einfluß des verbliebenen azimutalen Magnetfeldes stellt sich dann eine nicht-axialsymmetrische Form der MRI, die sogenannte azimutale MRI (AMRI), ein. AMRI ist aus astrophysikalischer Sicht besonders interessant, da sie nicht auf ein von außen vorgegebenes vertikales Magnetfeld angewiesen ist, sondern das erforderliche azimutale Magnetfeld gegebenenfalls durch einen gekoppelten MRI-Dynamo-Prozess selbst erzeugen kann. Die jüngsten AMRI-Experimente an PROMISE haben neben dem prinzipiellen Nachweis dieser Instabilität auch eine handfeste Überraschung geliefert: AMRI arbeitet in einem signifikant größerem Parameterbereich als ursprünglich erwartet. Inzwischen haben detaillierte numerische Simulationen die Ursache hierfür im verblüffend starken Effekt der durch die Stromzuführungen bedingten leichten Symmetriebrechung des angelegten Magnetfelds identifizieren können.

Eine ebenfalls nicht-axialsymmetrische Instabilität entsteht, wenn der axiale Strom nicht isoliert vom Fluid, sondern durch das Fluid hindurch fließt. Die dann auftretende Tayler-Instabilität (TI) wurde ebenfalls am HZDR nachgewiesen. Höchst interessante Perspektiven ergeben sich aus der möglichen Kombination von AMRI und TI. Neue theoretische Analysen haben gezeigt, dass schon bei einem geringen Anteil des Gesamtstromes im Fluid HMRI und AMRI durchaus in der Lage wären, auch Keplersche Rotationsprofile zu destabilisieren.

Im Zusammenhang mit diesem Effekt, sowie mit der oben erwähnten starken Wirkung einer leichten Symmetriebrechung des angelegten Feldes, hat PROMISE eine intensive Debatte über die Rolle von HMRI und AMRI im astrophysikalischen Kontext angeregt. Diskutiert wird insbesondere, ob das bisher auf elektrisch gut leitfähige Bereiche von Akkretionsscheiben beschränkte Anwendungsgebiet der Standard-MRI durch HMRI oder AMRI auch auf weitere Gebiete, insbesondere auf die sogenannten „dead zones" im Zentrum protoplanetarer Scheiben, ausgedehnt werden kann. In diesem Sinne haben die Bemühungen um die experimentelle Realisierung von MRI fruchtbare Rückwirkungen auf die astrophysikalische Forschung.

## 5 Zusammenfassung und Ausblick

Die ersten erfolgreichen Experimente in Riga und Karlsruhe waren ein Durchbruch bei der Erforschung des homogenen Dynamoeffekts. Das französische VKS-Experiment hat durch das Auftreten spontaner Feldumpolungen zu einem vertieften Verständnis des entsprechenden geophysikalischen Phänomens beigetragen. Während die Resultate des Rigaer und Karlsruher Experiments sehr gut mit den jeweiligen numerischen Vorhersagen übereinstimmten, haben gerade die überraschenden Ergebnisse des VKS-Dynamos weitere Forschungen initiiert. Eine ähnliche Entwicklung zeichnet sich in Hinblick auf die neuesten MRI-Experimente ab. Und genau hier wird es spannend: wenn Experimente unerwartete Resultate zu produzieren, die daraufhin von der Theorie erklärt werden müssen, woraus sich im besten Fall ganz neue Fragestellungen für den ursprünglichen Anwendungsbereich in Geo- und Astrophysik ergeben. In diesem Sinne kann man von den laufenden und zukünftigen Experimenten noch viele interessante Ergebnisse und Anregungen erwarten.